# A Performance Study of a Fast-Rate WLAN-Fingerprint Measurement Collection Method

Erick Schmidt, *Student Member, IEEE,* Misbahuddin A. Mohammed, David Akopian, *Senior Member, IEEE*

***Abstract*— Indoor positioning systems exploiting WLAN signal measurements such as Received Signal Strength (RSS) are gaining popularity due to high accuracy of the results. Sets of RSS and other measurements at designated locations from available WLAN access points (APs) are conventionally called fingerprints and retrieved from network cards at typically one Hz rate. Such measurement collection is needed for offline radio-map surveying stage which assigns fingerprints to locations, and for online navigation stage, when collected measurements are associated with the radio-map for positioning. As WLAN network is not originally designed for localization, the network cards occasionally miss the fingerprints, measurement fluctuations necessitate statistical signal processing, and surveying process is very time consuming. This paper describes a fast measurement collection approach that addresses the mentioned problems – higher probability of measurement acquisition, more data for statistical processing and faster surveying. The approach is further analyzed for practical setting applications.**

## I. Introduction

PEDESTRIAN navigation gained significant attention in the last decade due to potential applications in location based services (LBS) ranging from commercial to emergency uses. While outdoor navigation is achieved by existing global navigation satellite systems (GNSS) such as the Global Positioning System (GPS) [1], Galileo [2], and Beidou [3], many indoor areas are typically inaccessible for GPS signals due to structural blockages and severe multipath propagation effects. Such areas are often referred to as GPS-denied environments. Different from outdoors, there is no universal solution for indoor navigation at the moment. But there are techniques which achieve very good accuracy using measurements from indoor wireless signaling infrastructures such as commonly deployed Wireless Local Area Networks (WLANs or Wi-Fi) [4], [5]. There are several challenges though in WLAN-based positioning that should be addressed for broader deployment of such techniques. Conventional positioning techniques had been applied for WLAN environments including time-of-arrival ranging-trilateration [6]-[10], and angle-of-arrival triangulation [11], [12] approaches. But most current WLAN-based indoor positioning systems (IPS) rely on received signal strength (RSS) measurements for localization techniques such as fingerprinting-based [13]-[16], and propagation modelling-based [17] approaches, among others.

The fingerprinting-based IPS approach is prevalent due to proven high-accuracy of the results [15], [16]. Different from conventional trilateration and triangulation methods, the fingerprinting gains from indoor multipath propagation in cluttered environments. These techniques collect wireless signatures for a grid of locations and design a radio-map database which records fingerprints at known grid locations for the area of interest. This is accomplished in offline mode. The online positioning process includes fingerprint measurements capture for the unknown location followed by matching the fingerprint to the closest radio-map entry [15]. Most of literature focuses on location estimation approaches given online measurements and pre-collected offline radio-maps. Depending on the mode of estimation, WLAN-based positioning methods are categorized as deterministic or probabilistic [4], [13], [15], machine learning [18]-[20], among others.

At the same time, an important process of radio-map measurement collection received less attention [5]. Generally, these surveying exercises are manpower and time expensive as the performance of radio-map construction is critical [21]. One should note also that WLAN infrastructure is not designed for positioning as fingerprinting measurements do not provide guaranteed performance for this same reason. One aspect of this problem is the selection of measurements from only reliable WLAN access points (APs) [22]-[24]. The filtering of faulty measurements, or *outliers*, from the available APs is another approach to improve the reliability of the measurements [25]-[27]. Distortion impact of various *outlier* types on WLAN-based fingerprint positioning is reported in [28], [29].

There are many ways to survey an area with wireless technology such as proprietary WLAN surveying equipment [30]-[32]. But many times, existing commercial WLAN network interface controllers (NICs) in laptops and

E. Schmidt is with the University of Texas at San Antonio, San Antonio, TX 78249 USA (e-mail: erickschmidtt@gmail.com).

M. A. Mohammed was with the University of Texas at San Antonio, San Antonio, TX 78249 USA (e-mail: misbahuddin.aamer@gmail.com).

D. Akopian is with the University of Texas at San Antonio, San Antonio, TX 78249 USA (email: david.akopian@utsa.edu).

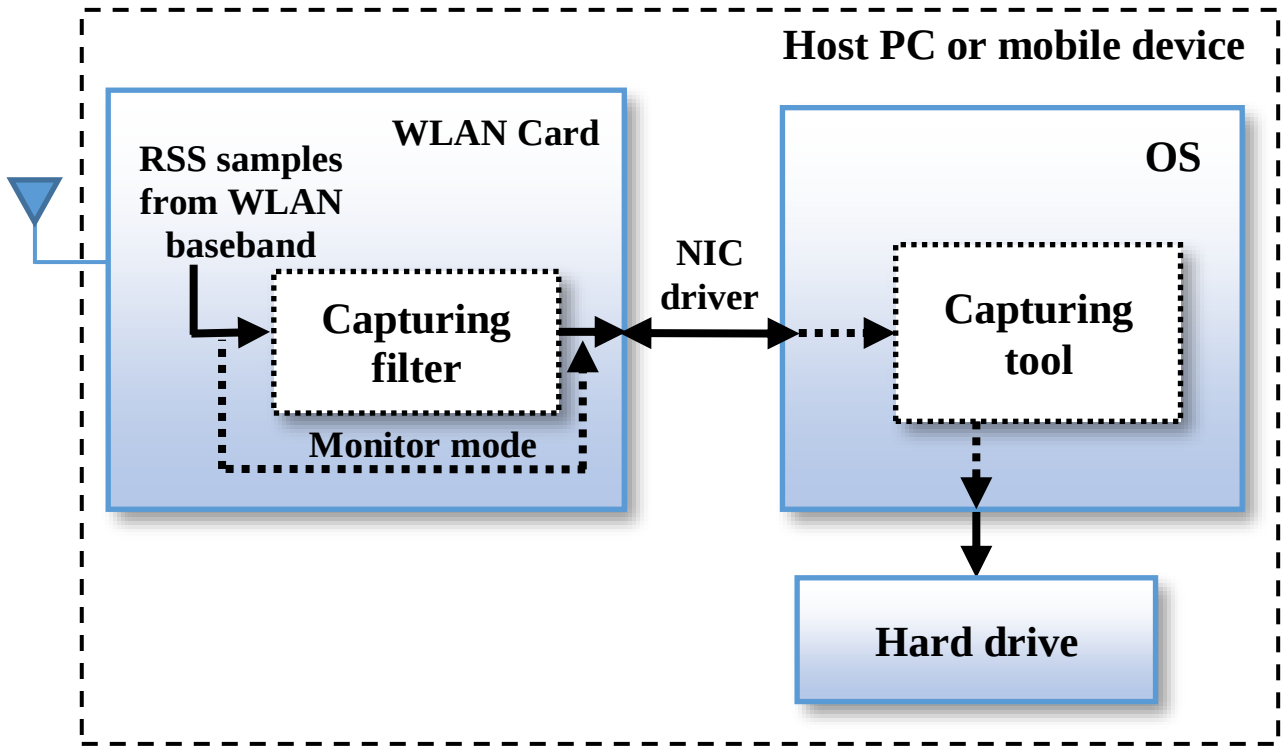

Fig. 1. Interface diagram showing interaction between the RSS samples, WLAN card, capturing filter, NIC driver, capturing tool, and hard drive.

smartphones are the affordable option since the former can be expensive.

When extracting RSS data from mobile devices and commercial NICs, specialized data collecting software have measurement rate limitations since output samples depend on many factors such as WLAN card vendor, NIC driver, operating system (OS) compatibility, among other considerations [33]. One Hz measurement collection rate is common [5], [34], [35] which magnifies the importance of measurement quality. First, the radio-map collection becomes very time consuming, as advanced probabilistic and machine-learning location estimation techniques require hundreds of measurements at each location for statistically sound results. To some extent this issue can be addressed using coarser-grid radio-map and applying interpolation techniques to preserve positioning accuracy [36], [37]. This typically results in accuracy degradations. Second, due to high rate of outliers including missing measurements caused by transient effects in network cards [34], advanced outlier filtering algorithms should be applied, but their performance depends on the number of available APs [27], [37].

This paper addresses the above-mentioned issues from another perspective. A method of extracting measurements from a single AP at about ten times faster rate compared to conventional 1 Hz rate is proposed, followed by an assessment of measurement rate and measurement availability including a methodology for testing proposed fast-rate measurement extraction mode vs. conventional methods. High rate measurement capture increases the probability of measurement availability within conventional one second analysis period. Additionally, it accelerates proportionally radio-map surveying time through faster sample collection. In a one second period, the probability of availability of an RSS sample is drastically increased with the proposed method, thus nullifying so-called outliers, specifically missed packets which degrade the performance of WLAN positioning as studied in detail in literature [28], [29], [34], [35], [38]. These outliers come either from faulty APs, interferences, or "transient effects" from WLAN NIC vendors. This paper focuses mostly on missing measurements resulting from different phenomena such as transient effects from NICs, unintentional interference, and faulty APs.

A faster measurement rate is achieved by exploiting so-called *monitor* mode of the network card. Fig. 1 shows the interfaces of the RSS samples with the mobile device's NIC, NIC driver, data collecting tool, and hard drive used when capturing packets inside a host PC. The method employs an open source networking tool that switches the NIC to a *monitor* mode. RSS samples are then collected directly from the NIC driver output at high measurement rates.

The paper is organized as follows. Section II describes two data collection modes: conventional and proposed fast-rate method. Section III proposes testing methodology to assess overall sample availability for both modes. Section IV presents evaluation results for proposed tests and data collection modes. Finally, future work is discussed in Section V, and conclusion remarks are made in Section VI.

## II. Data Collection Modes

Due to the nature of Wi-Fi's functionality as a wireless communication environment, the conventional WLAN cards are exposed to many traffic patterns at a given time. Because of this intense traffic, these cards are typically configured to communicate only relevant information with their host devices using filters and other packet manipulations, thus prioritizing relevant user-dedicated packets for saving power and computation resources. There are data, management, and control packets being broadcasted in the wireless medium (WM) from numerous APs, which are filtered in the default modes of operation for most commercial cards. The cards also limit packet capturing rates in the conventional modes of operation.

Two filters are of interest for the following discussion, which are used inside WLAN cards in user-dedicated mode of operation: SSID filter which prioritizes to currently connected wireless network, and WLAN channel filter, which only listens for packets on current channel [33]. A capturing tool such as Wireshark [39] can be used to collect packets in this mode of operation (which will be referred to as *normal* mode for testing scenarios in this paper). For this mode, the capturing chain can be seen as follows: WLAN card → capturing filter → capturing tool → hard drive [33] (see Fig. 1). The prevailing majority of the reported WLAN-fingerprint measurement collection techniques exploit this type of data collection.

To overcome WLAN measurement data and rate acquisition limitations, this paper suggests collecting data in so-called *monitor* mode, by bypassing the filters. The NIC driver works as a mere interface between the WLAN card, the OS, and the capturing tool found in the host PC. In *monitor* mode, all broadcast packets for all visible access points (APs) and for all compatible WLAN channels are available for capturing. This can be accomplished by, e.g. *Aircrack* [40], an open-source tool that modifies operation mode of the NIC driver to *monitor* mode and is primarily used in Linux OS. The capturing chain is now: WLAN card → capturing tool → hard drive (see Fig. 1). This mode of operation removes previously mentioned filtering restrictions of the card for faster capturing. Depending on the vendor and/or driver, this tool can also remove user-dedicated operation. More information on how to install *Aircrack* on compatible NICs is found on the website [41]. This tool sets the WLAN card to *monitor* mode through the NIC driver. While the *monitor* mode allows to overcome limitations of WLAN fingerprint measurement collection, there were no reports on the performance of such measurements. In the following, the paper investigates both *normal* and *monitor* modes of WLAN card operations for measurement collection in representative testing environments at the University of Texas at San Antonio (UTSA), as well as at a nearby student apartment complex.

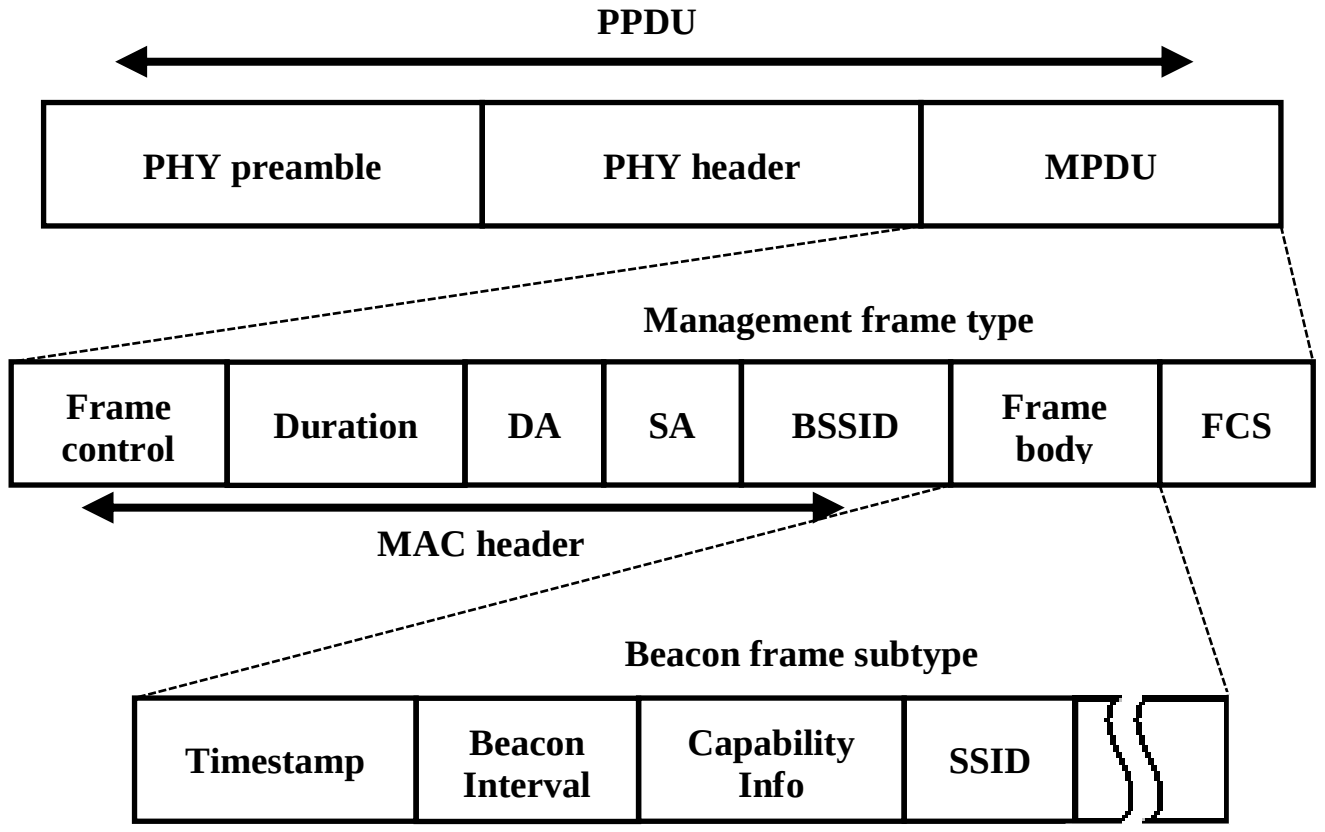

Fig. 2. PPDU and beacon frame packet description fields based on 802.11 standard.

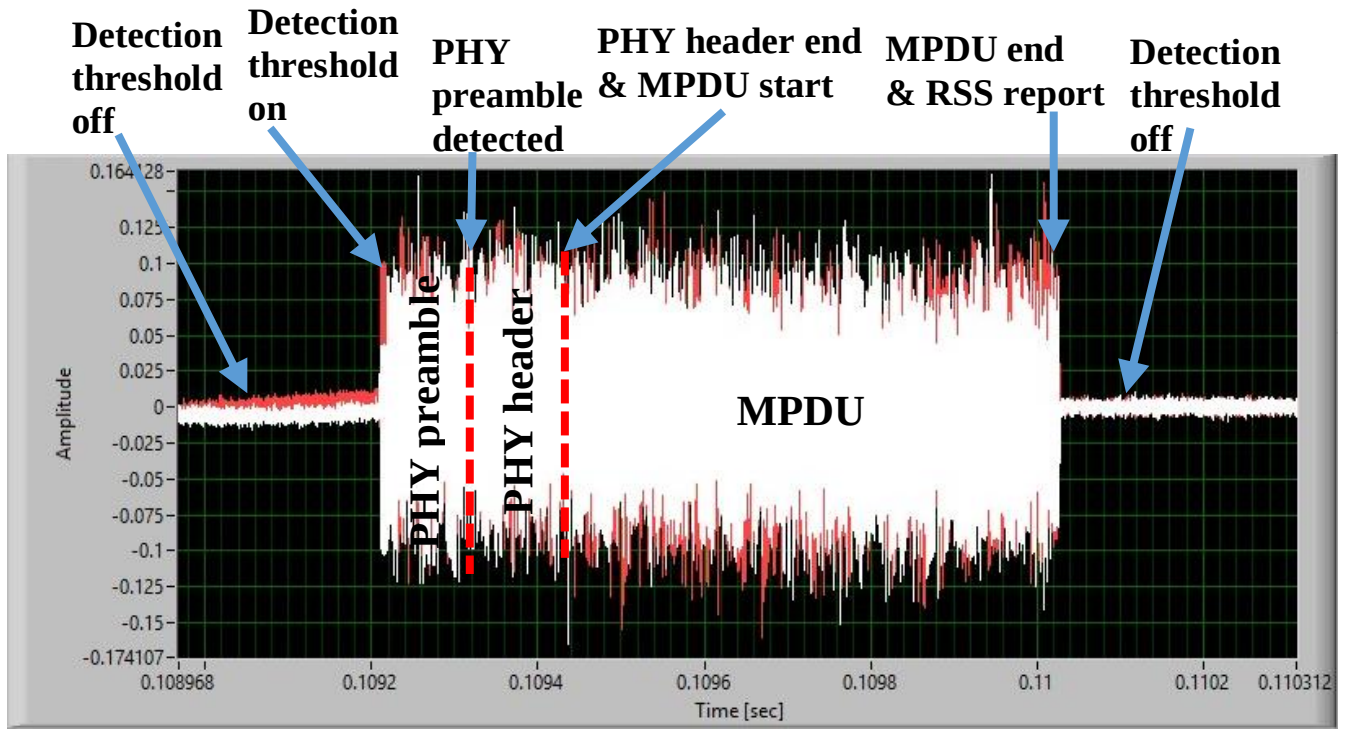

Fig. 3. WLAN signal recording with NI-USRP and LabVIEW showing (approximate) lengths of a WLAN frame and RSS computation.

### A. WLAN beacon frame structure

RSS samples are extracted from specific captured packets named physical (PHY) layer packet data units (PPDUs). PPDUs contains a medium access control (MAC) beacon frame, which is broadcasted from each AP. This is a broadcast management frame transmitted by all APs in their own time slots. These packets are used by the receiving host NIC to locally compute and report RSS since it contains a known training sequence specified on the 802.11 standard [42] for detecting and processing a WLAN PPDU. This training sequence is found on the PHY preamble. A PHY header is also commonly used to demodulate the MAC packet data unit (MPDU) where the rest of the beacon frame information resides. The WLAN card will report RSS every time it successfully decodes the full beacon frame. Fig. 2 shows the arrangement of the PPDU and the MPDU beacon frame fields, where some of these fields are used for relating the reported RSS value from that packet with the specific AP. This is done by extracting information such as service set identifier (SSID) and MAC address relevant to that AP. This information is fundamental to fingerprinting-based IPS. Technically speaking, RSS is measured by the WLAN card after capturing a broadcast packet. The packet that is being captured and that is of interest is a management type with beacon frame as subtype. These categories are part of the 802.11 standard. Once the packet has been completely received and demodulated, the WLAN card reports the RSS value at the end of the packet, thus associating the reported RSS reading with that packet. Fig. 3 shows a real signal recorded with National Instruments USRP instrumentation [43] to mark the entire process occurring internally in the WLAN card. It shows approximate boundaries of WLAN beacon frame fields such as PHY preamble, PHY header, and the MPDU. WLAN cards have a power level threshold used to sense a packet in the WM. Once this threshold is reached the WLAN card begins to search for a potential WLAN frame, i.e. by correlating with the known PHY preamble, as defined in [42]. After this, the PHY header describes what is contained in the MPDU and how to demodulate it. Once the MPDU is finally demodulated and information extracted from the fields, the WLAN card reports the RSS to the WLAN driver with its associated information such as SSID, MAC address, etc. The frequency of this report depends on the mode of operation (*normal* or *monitor*).

### B. WLAN beacon frame sampling period

The beacon frame from a single AP can be transmitted up to every 102.4 milliseconds which is defined as 100 time units (TU) by the 802.11 standard [42] (1 TU is 1.024 milliseconds). This beacon frame periodicity is limited by the standard itself. Therefore, assuming direct measurement extraction from the NIC card in *monitor* mode, one can theoretically acquire up to 9.77 RSS measurements per second per AP.

In *normal* mode, the RSS measurements acquisition is typically reported at 1 Hz rates [5], [34], [35]. Our experiments demonstrate measurement rate interval of 1,000 TUs which is every 1024 milliseconds (roughly a second). This translates to theoretical measurement rate of 0.977 RSS packets per second in *normal* capturing mode. Both *normal* and *monitor* theoretical rates will be used as a reference in our tests defined in Section IV.

As it was mentioned, the assessment of *monitor* mode capturing capabilities attracts the attention because of improved rate of measurement collection. This eventually translates to faster surveying times and almost guaranteed availability of RSS samples captured in one second as will be studied in the next sections. Since now there are more samples to choose from one can use extra measurements for averaging, selection, etc., as conventionally exercised in WLAN positioning methods.

## III. Testing Methodologies

This section presents testing scenarios which approach realistic and representative surveying situations, to assess the proposed *monitor* mode capturing method versus *normal* mode capturing. The selected testing environments are student apartments and university environments, which are usual for WLAN positioning research.

In general, testing methodologies in *normal* mode are applied in Windows OS, and *monitor* mode is used in Linux OS, respectively. The testing equipment used for indoor surveying was an ASUS X555LA series laptop with a 4th generation Intel Core i5 at 1.7 GHz, 8GB of RAM, and Windows 10 Home, as well as Linux Ubuntu 16.04 LTS in dual-boot mode. This same laptop comes with an integrated WLAN NIC Atheros model AR9485 with 802.11b/g/n capabilities. A Ralink RT3070 USB dongle with same Wi-Fi capabilities was used as a second choice for testing. Both cards are compatible with *Aircrack* in Linux. The capturing tools used were Wireshark version 2.0.12 [39] for Linux OS, and Acrylic Wi-Fi Professional v3.2.6269.20454 [44] for Windows OS.

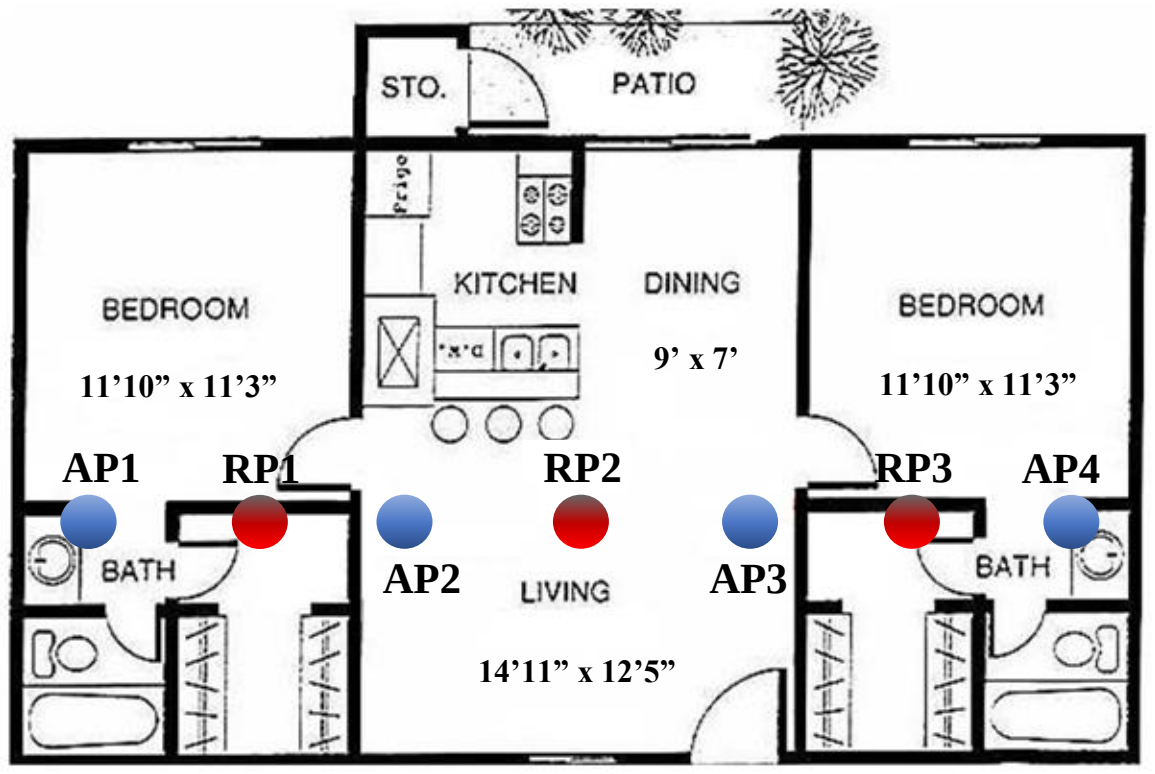

Fig. 4. Apartment floorplan AP/RP configuration for traffic test.

Three assessment testing campaigns were conducted to observe capturing capabilities and measurement data collection rates, using both WLAN cards and both operating modes (*normal* vs *monitor*): *traffic* test, *AP signal strength (distance)* test, and *vendor CPU load* test, which are described next. A fourth test, i.e. *probability of capture* test, was derived from the *AP signal strength (distance)* test as well. For improved analysis, for each test scenario, a total of 200 seconds capturing time, performed 10 times, was collected. This gives a total of 33.33 minutes worth of data capturing per test scenario.

## A. Traffic test

It is hypothesized that WLAN signal interferences may negatively impact measurement capture rate. For this reason, a testing is performed in a crowded WLAN deployment environment. The *traffic* test was performed in a student apartment near the university, where 4 WLAN routers were placed throughout the apartment and three reference points (RP) were used for measurement data collection. Also, around 5 people were at the apartment during the tests and APs were being used normally. Also, the apartment itself is prominently constructed of wood materials, therefore as much as 20 APs were also visible from neighbor apartments. Fig. 4 shows the floor plan and testing configuration for said APs and RPs. This test was performed on Atheros NIC only for both *normal* and *monitor* modes.

## B. AP signal strength (distance) test

This test assesses impact of distance from the AP in terms of received signal strength on measurement capturing performance rate of the WLAN card. Typically, less measurements are captured in weaker signal conditions, and the tests will demonstrate the improvement of measurement availability due to *monitor* mode exploitation. One AP was placed in a hallway of the 2nd floor of the AET building at UTSA. Fig 5. shows a floorplan of the test location. Two AP signal strengths were measured with averaged RSS values of -60 dBm and -80 dBm, which are considered strong and marginal signal levels, respectively. This test was performed on Atheros NIC for *normal* and *monitor* modes.

## C. Vendor CPU load test

In this test, it is hypothesized that computational stress loads on host CPU can results in transient impacts on measurement capture capabilities. For this test, a CPU stress tool *Heavy Load v3.4* [45] was used in Windows environment, and *stress-ng* [46] was used in Linux. Without the loss of generality, test scenarios of 50% and 80% CPU load were assessed for both *normal* mode and *monitor* mode, for both Atheros and Ralink NICs. The test was performed only in marginal WLAN signal conditions (-80dBm) to compare against a real-world worst-case surveying situation in which the mobile device is multitasking while collecting RSS packets in the background, such as in [14], for either offline surveying or online navigation. This test was performed in the same hallway location as the distance test (see Fig. 5), with similar conditions.

## D. Probability of capture test

An overall assessment of the capabilities of capturing measurement data at fast-rate *monitor* mode versus *normal* mode is conducted as well. A simple way to compare both capturing modes is to define probability of capturing a packet in a one-second interval, which is a reported conventional measurement rate [5], [34], [35]. Although *monitor* mode is capable of capturing at faster rates, the comparison uses as a base the *normal* mode of operation since we observe and leverage the availability of samples at a given period for the proposed capturing method. This specific test is obtained directly from the *AP signal strength (distance)* test at weak signal conditions for both modes.

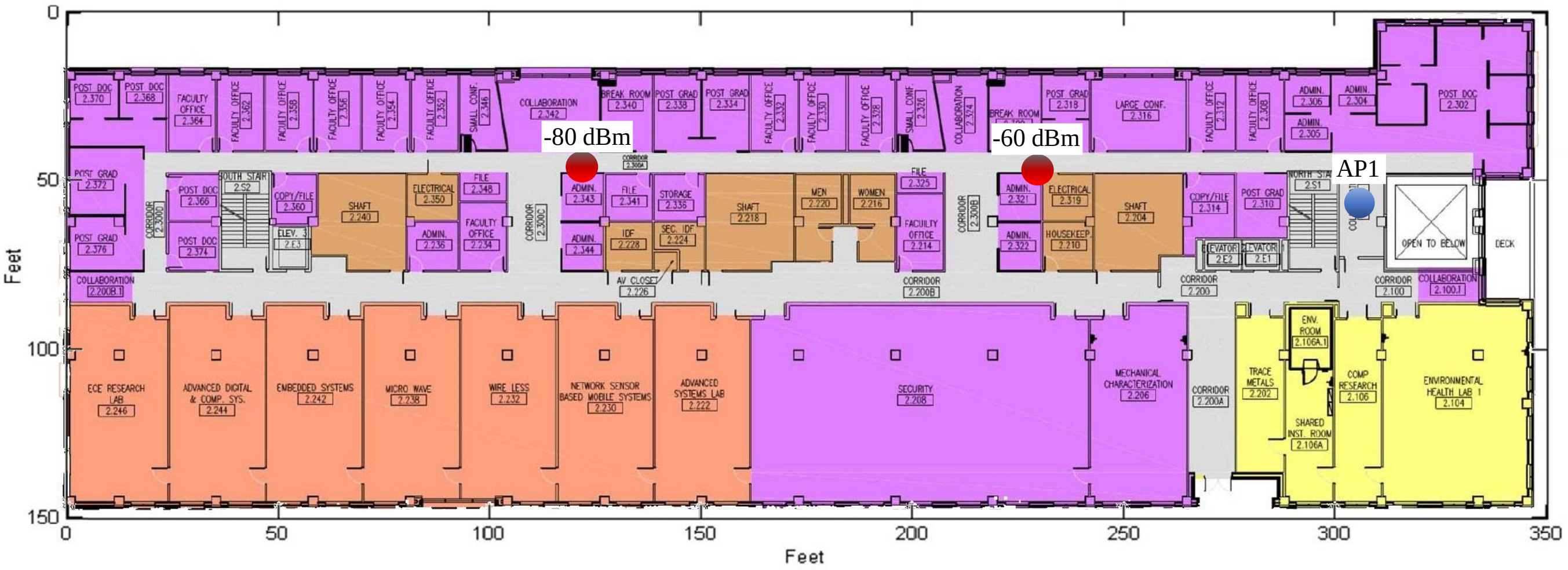

Fig. 5. UTSA's AET building 2nd floor hallway for AP signal strength (distance) and vendor CPU load tests.

TABLE I
TRAFFIC TEST RESULTS FOR NORMAL AND MONITOR MODE

| | Normal mode | | Monitor Mode | |
|---|---|---|---|---|
| | Avg. meas. rate, packets per sec. | Miss-rate (%) *compared to theoretical | Avg. meas. rate, packets per sec. | Miss-rate (%) *compared to theoretical |
| AP1 | 0.90 | 7.50% | 4.76 | 51.25% |
| AP2 | 0.91 | 7.09% | 7.40 | 24.22% |
| AP3 | 0.91 | 6.93% | 7.42 | 24.06% |
| AP4 | 0.91 | 7.03% | 8.29 | 15.10% |

## IV. RESULTS

This section presents assessment results on RSS measurement capture rates. Histogram plots related to WLAN beacon packet time interval show the spread of packet arrival delays. This way one can observe different phenomena on packet arrival delays spread on these Histogram plots, for the different scenarios. The average number of received measurement packets per second is evaluated as well as the percentage of missing measurements (miss-rate) in packets per second vs. the theoretical maximum measurement rate value in packets per second for both *normal* and *monitor* mode. These theoretical values are 0.977 and 9.77 samples per second, for *normal* and *monitor* mode respectively, and are defined as the theoretical maximum achievable samples per second for each capturing mode (see Section II). An important observation factor in all results is the availability of at least one measurement in the proposed *monitor* mode.

### A. *Traffic test results*

Table I shows results for traffic test. The average testing time from all 10 runs was used to compute the rate of received measurement packets per second, and percentage of missing packets, referred to as *miss-rate*, from theoretical maximum measurement rate, in packets per second. We also used capture rates from all three reference points (see Fig. 4) to obtain improved averaged values for each AP. By looking at the miss-rate for both modes, one can observe that *normal* mode is less affected in high traffic environments. The WLAN NIC filtering aids in high traffic situations by ignoring most of the traffic in the WM, therefore low packet loss is achieved. An average of 0.90 packets per second was attained for *normal* mode and around 7% of missing packets was observed, compared to the theoretical maximum value. On the other hand, *monitor* mode is capturing all the available broadcast traffic and thus many packets can be lost due to a busy WM and possible multi-path effects. The measurement rate from AP1 suffered the most, losing 51.25% of measurement rate, and achieving an average of 4.76 packets per second measurement capture rate. Despite such losses, the *monitor* mode in AP1 still captured at 5 times faster rate when compared to *normal* mode due to an unfiltered

TABLE II
AP SIGNAL STRENGTH (DISTANCE) TEST RESULTS FOR NORMAL AND MONITOR MODE

| | Normal mode | | Monitor mode | |
|---|---|---|---|---|
| | Avg. meas. rate, packets per sec. | Miss-rate (%) *compared to theoretical | Avg. meas. rate, packets per sec. | Miss-rate (%) *compared to theoretical |
| -60 dBm (strong) | 0.91 | 7.04% | 9.68 | 0.86% |
| -80 dBm (weak) | 0.57 | 41.67% | 9.22 | 5.63% |

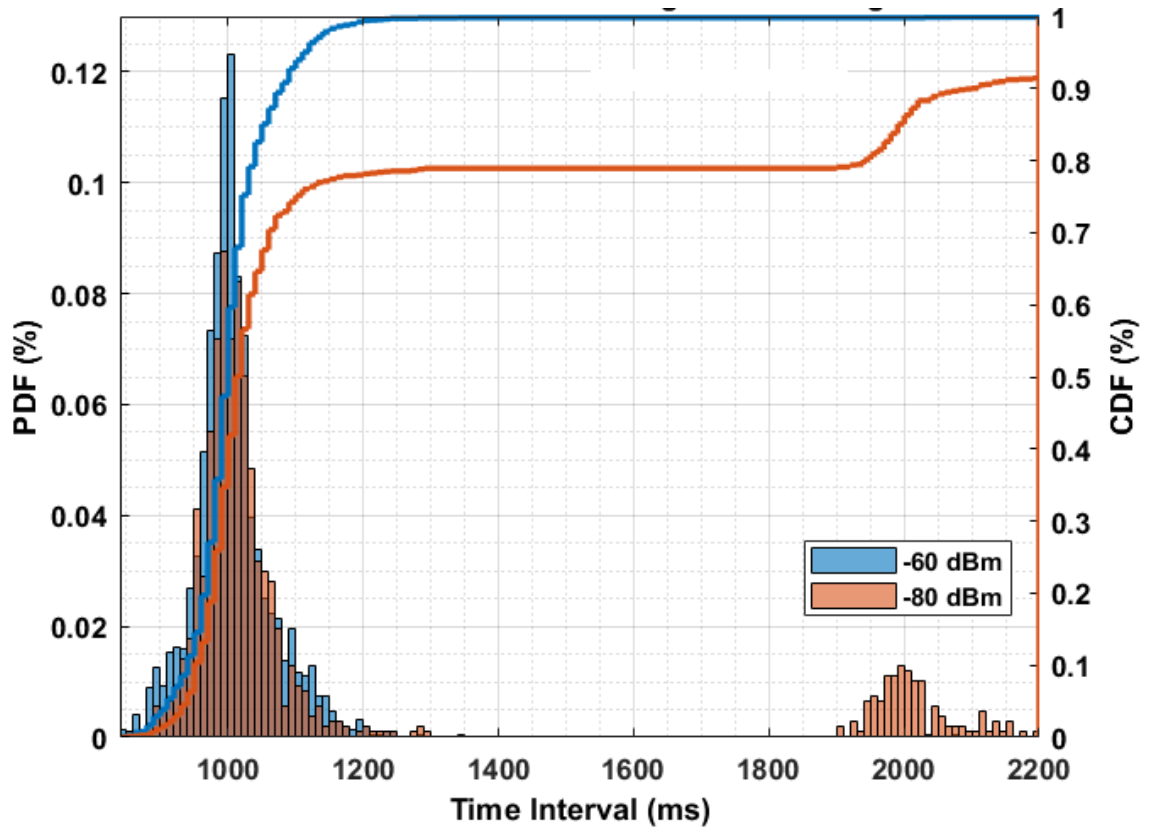

Fig. 6. *AP signal strength (distance)* test packet arrival delay histogram in *normal* capture mode for strong signal conditions (-60 dBm) and weak signal conditions (-80 dBm).

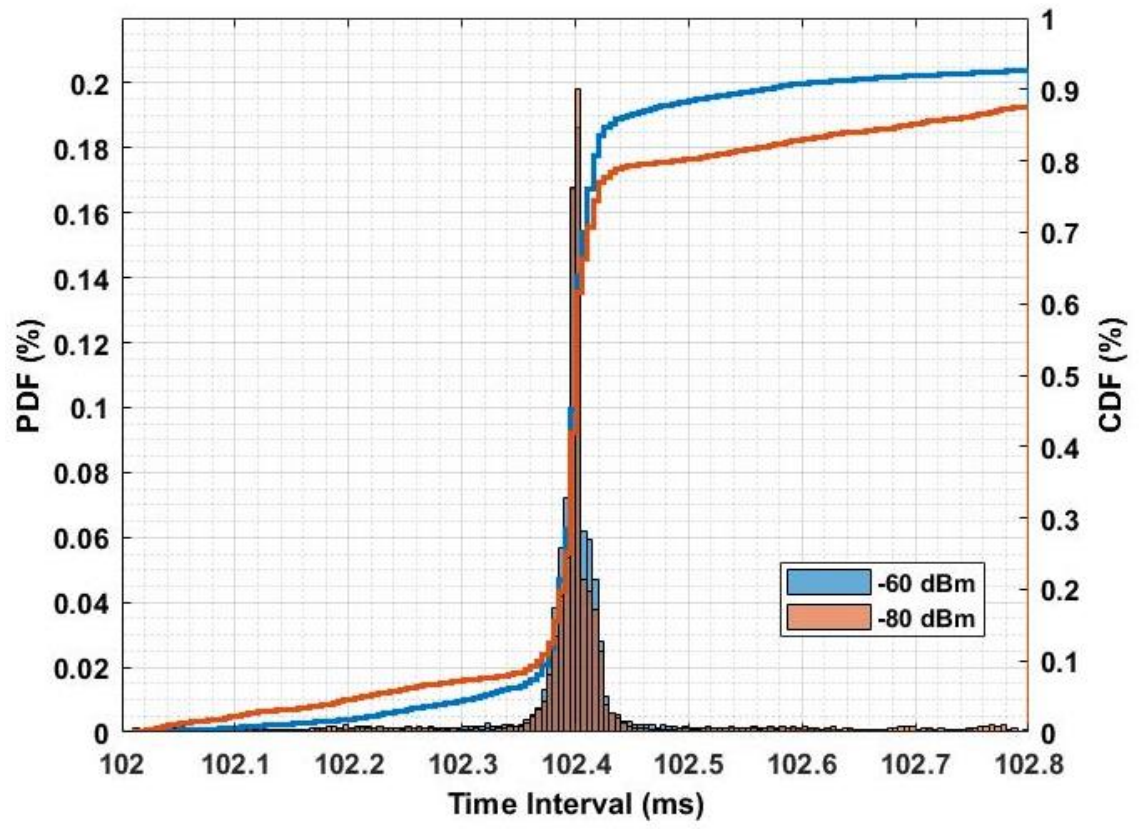

Fig. 7. *AP signal strength (distance)* test packet arrival delay histogram in *monitor* capture mode for strong signal conditions (-60 dBm) and weak signal conditions (-80 dBm).

higher rate of operation, therefore improving sample availability.

### B. *AP signal strength (distance) test results*

Table II shows the total number of packets received for both strong and weak AP signal strengths (distances), for *normal* and *monitor* mode respectively. An important loss of packets can be seen between the strong and weak signal distances for *normal* mode, showing a 49.2% difference in average measurement rates for the observation duration. The strong signal distance shows a result close to the theoretical maximum, similar to traffic test results for *normal* mode: 0.91 packets per second, with a miss-rate of 7%. One can also see a significant drop in received packets rate for weak signals: 0.57 packets per second translating to a miss-rate of 41%. This is because weak signal packets are not captured by the NIC receiver when filtering is used in *normal* mode.

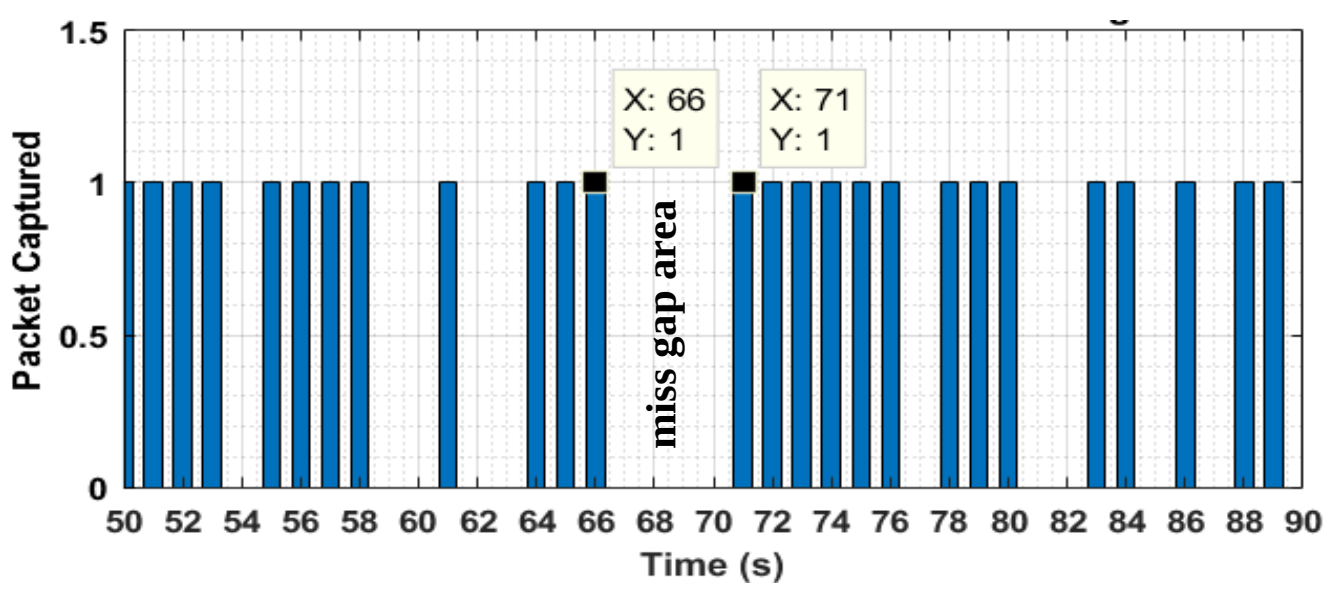

Fig. 8. Packet miss phenomena observed during *AP signal strength (distance)* test in *normal* mode for weak signal conditions. Between seconds 66 and 71, a miss gap area is observed.

Another way to observe these phenomena is to check packet delivery delays on histograms. The packets which are not delivered are being retransmitted until the next time slot, which retranslates the issue of missed packets to packet delays. Fig. 6 shows packet arrival delay histogram plots for both strong and weak AP signals to illustrate such delays. The way this histogram plot was generated was by computing the difference in the timestamps between consecutively captured samples, therefore obtaining values such as, e.g., 1000 ms, 1003 ms, 1004 ms, 2000 ms, etc. for *normal* mode. Similarly, Fig. 7 displays packet arrival delay histogram for *monitor* mode, showing minimal influence in miss-rate, as seen in Table II.

*1) Packet miss phenomena*

Since for *normal* mode, packets are captured at roughly 1 Hz, a packet miss can impact highly on measurement rate. If the packet was not received near 1 second time interval, it will not then arrive until the next second interval (time slot) for *normal* mode. Therefore, this packet miss phenomena can highly impact overall measurement sample availability in weak signal conditions. Fig. 8 shows packet misses in *normal* mode for *AP signal strength (distance)* test in weak signal conditions vs. time, for seconds 50 to 90 for a given recording out of 10, for a single AP. For each plotted second in the x-axis, a bar represents a captured packet, to illustrate how packets were captured during time of capture. The most predominant packet misses occur between seconds 66 to 71, having 4 consecutive seconds without capturing a packet. For certain IPS, continuous RSS packet collecting for a single AP is crucial for positioning determination [28], [29]. This observed phenomena directed attention to a necessity of *vendor CPU load* tests to compare this same weak-signal condition added with a CPU stress. These two conditions combined will highlight hypothesized transient effects of the NIC or mobile device [34], thus generating packet misses for said phenomena. Thus “transient effects” are evaluated by said test.

## C. Vendor CPU load test results

As seen in the *AP signal strength (distance)* test results, a hypothesized impact on measurement rate is expected in this test because of the added stress on CPU. Table III shows vendor comparison CPU load tests at weak-signal distance conditions for *normal* mode. Overall, Ralink vendor underperformed compared to Atheros’ measurement rate while both experienced a relative difference in packet measure rate of roughly 11% when stress is applied in *normal* mode. Ralink showed an average measurement rate of 0.69 packets per second on high CPU load, which translates to 29.45% miss-rate from theoretical maximum measurement rate. As mentioned in Section IV-B.1, packet misses due to transient effects (see Fig. 8) are analyzed in v*endor CPU load* test. Fig. 9 shows an example for Atheros *normal* mode packet delay histogram chart for both 50% and 80% CPU loads for said test, similar trend is observed in Ralink NIC results. A decrease in the 1 second main lobe vicinity can be seen spread towards 2 second delays when CPU stress is increased, thus proving a possible transient effect on the NIC.

TABLE III
VENDOR CPU LOAD TEST RESULTS FOR NORMAL MODE

| | Atheros NIC | | Ralink NIC | |
|---|---|---|---|---|
| CPU load | Avg. meas. rate, packets per second | Miss-rate (%) *compared to theoretical | Avg. meas. rate, packets per second | Miss-rate (%) *compared to theoretical |
| 50% load | 0.96 | 2.17% | 0.76 | 21.92% |
| 80% load | 0.86 | 11.86% | 0.69 | 29.45% |

TABLE IV
VENDOR CPU LOAD TEST RESULTS FOR MONITOR MODE

| | Atheros NIC | | Ralink NIC | |
|---|---|---|---|---|
| CPU load | Avg. meas. rate, packets per second | Miss-rate (%) *compared to theoretical | Avg. meas. rate, packets per second | Miss-rate (%) *compared to theoretical |
| 50% load | 8.76 | 10.28% | 8.19 | 16.13% |
| 80% load | 7.80 | 20.17% | 5.05 | 48.30% |

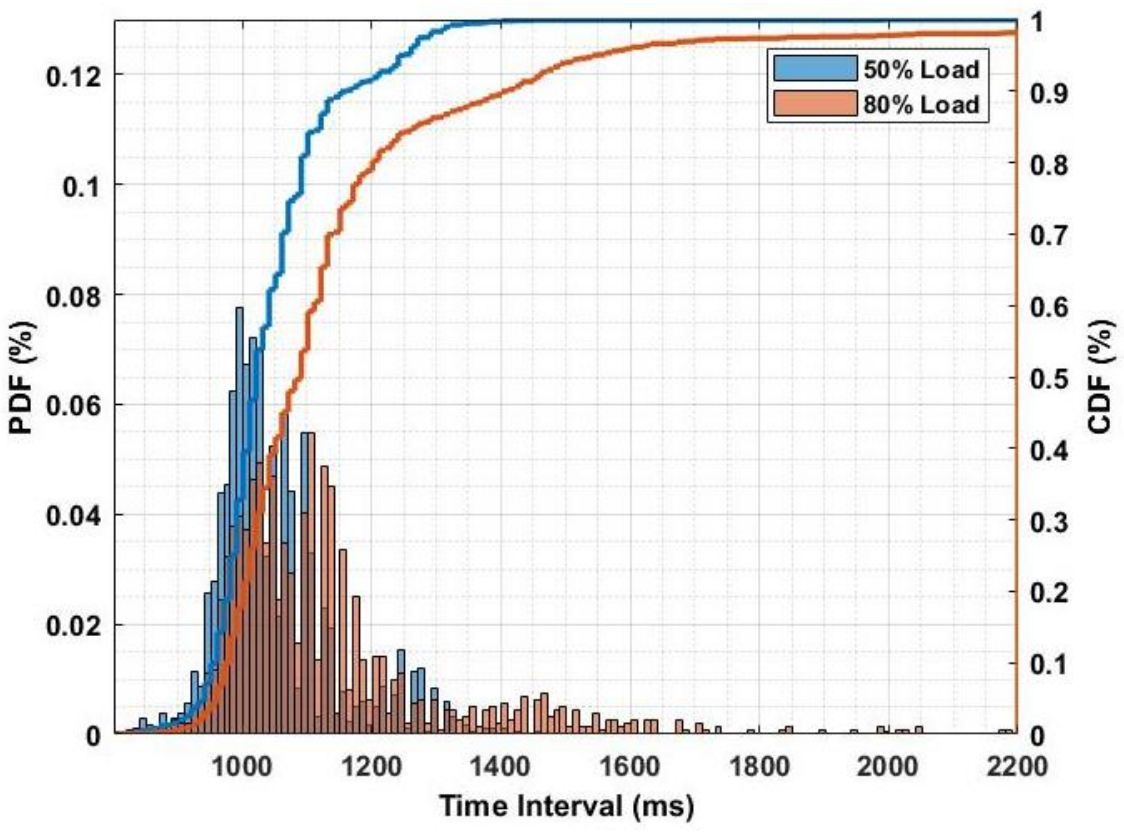

Fig. 9. Atheros *vendor CPU load (stress) test* packet arrival delay histogram in *normal* capture mode for 50% and 80% CPU load at weak signal conditions.

Results for *monitor* mode for both vendors can be seen in Table IV. While Atheros NIC mantains same difference percentage for packet measure rate in *normal* and *monitor* mode with different loads, Ralink shows weaker performance of 47.41% difference in packet measure rate between both loads and up to 48.30% miss-rate from theoretical maximum. Fig. 10 shows packet arrival delay histogram for Ralink under CPU stress, providing visual aid to weaker performance observation seen in Table IV when stress is increased. This transient effect is seen in *monitor* mode on Ralink NIC in Fig. 10.

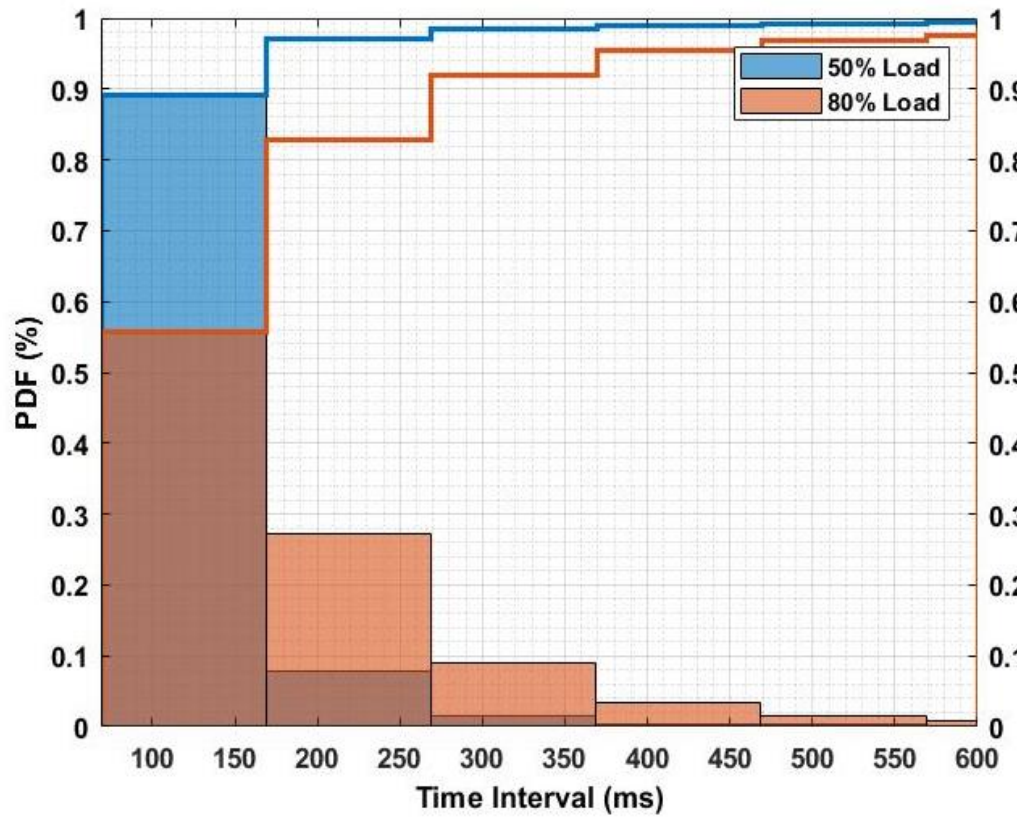

Fig. 10. Ralink *vendor CPU load (stress) test* packet arrival delay histogram in *monitor* capture mode for 50% and 80% CPU load at weak signal conditions.

TABLE V
PROBABILITY OF CAPTURE FOR NORMAL AND MONITOR MODE (-80DBM)

| | Monitor mode probability of capture at 1 Hz | Normal mode probability of capture at 1 Hz | Normal mode probability of capture at 2 Hz |
|---|---|---|---|
| Probability of packet capture (%) | 100% | 35.1% | 85.3% |

### D. Probability of capture test results

This section provides an integrative assessment of measurement availability in *normal* and *monitor* modes considering one or two second intervals. As the measurement rate of *monitor* mode is significantly higher, the probability of availability of at least one packet is very high. Table V summarizes results for one of the case scenarios. At 1 Hz, *monitor* mode showed a probability of packet capture of 100% of one or more measurement capture for *AP signal strength (distance)* test at weak-signal conditions (-80 dBm) as opposed to *normal* mode which showed a very low 35.1%. A second statistic is shown in Table V for *probability of capture* in a 2 seconds period in which *normal* mode shows 85.3% probability of a packet capture. This interprets numerous packet misses showing until the following second or time slot, thus impacting measurement availability when in *normal* mode.

## V. FUTURE WORK

The proposed methodology for testing leaves space for numerous scenarios as future work. As the presented testing scenarios were representative, it is desirable to conduct thorough testing in diverse environments for practical deployments. It is also common to investigate these phenomena across different NIC vendors as heterogeneous deployment studies [47], [48]. The open-source *monitor* mode tool is constantly updated to accommodate more number of NIC vendors but is certainly not available for all cards. This and future studies on the usefulness of *monitor* mode will motivate broader deployment of this feature, especially its applicability for mobile devices such as smartphones.

## VI. CONCLUSION

This paper discusses a performance study of sampling rates of RSS measurements on conventional samplers (i.e. 1 Hz) vs a proposed fast-rate method. The proposed fast-rate method almost guarantees availability of a sample on a given second, therefore minimizing common problem of missing measurements. It exploits an open-source tool which enables a so-called *monitor* mode of NIC drivers in common embedded WLAN-cards. This mode increases measurement capture rate almost tenfold. As the measurement capture is a random process, the paper analyzed the performance of the proposed instrumentation for different scenarios. The testing scenarios do not claim commonality, but they are representative of a university environment. At the same time, the introduced testing methodology on proposed capture modes is scalable for other possible scenarios. It is shown that *monitor* mode is significantly more reliable in capturing measurements compared to the commonly used 1 Hz positioning rate. It is also shown that the *monitor* mode is sensitive to interference as seen in the *traffic* test, but nevertheless provides higher measurement rate compared to the commonly used *normal* mode. It is observed that measurement rate depends on transient effects of WLAN cards and is also sensitive to the CPU loads of host devices. At measurement rates of 8-9 measurement packets per second, compared to commonly used 1 packet per second, the measurement availability in a one-second time is almost guaranteed. This benefits with man-hour surveying efforts when constructing radio-maps and improving the indoor positioning performance by minimizing missing measurements [28], [34]. One should note that not all WLAN cards can support typical data communication functions while in *monitor* mode.

## REFERENCES

[1] P. Misra and P. Enge, *Global Positioning System: Signals, Measurements, and Performance*, 2nd ed., Lincoln, MA: Ganga-Jamuna Press, 2006.
[2] “Galileo” European Satellite Navigation System, [Online]. Available: http://www.gsa.europa.eu/galileo/why-galileo. [Accessed May 2017].
[3] “Beidou” Chinese Satellite Navigation System [Online]. Available: http://en.beidou.gov.cn/. [Accessed May 2017].
[4] A. Khalajmehrabadi, N. Gatsis, D. Akopian, “Modern WLAN fingerprinting indoor positioning methods and deployment challenges,” *Commun. Surveys Tuts.*, vol. 19, no. 3, 2017.
[5] S. H. Jung, B.-C. Moon and D. Han, "Performance Evaluation of Radio Map Construction Methods for Wi-Fi Positioning Systems," *IEEE Trans. Intell. Transp. Syst.*, vol. 18, no. 4, pp. 880-889, April 2017.
[6] M. E. Rusli, M. Ali, N. Jamil and M. M. Din, "An Improved Indoor Positioning Algorithm Based on RSSI-Trilateration technique for Internet of Things (IoT)," in *Computer and Communication Engineering (ICCCE), 2016 International Conference on*, Kuala Lumpur, Malaysia, 2016.
[7] S. He and S.-H. G. Chan, "INTRI: Contour-Based Trilateration for Indoor Fingerprint-Based Localization," *IEEE Trans. Mobile Comput.*, vol. 16, no. 6, pp. 1676 - 1690, June 2017.
[8] X. Li and K. Pahlavan, “Super-resolution TOA estimation with diversity for indoor geolocation,” *IEEE Trans. Wireless Commun.*, vol. 3, no. 1, pp. 224–234, Jan. 2004.
[9] C.-R. Comsa, J. Luo, A. Haimovich, and S. Schwartz, “Wireless localization using time difference of arrival in narrow-band multipath systems,” in *International Symposium on Signals, Circuits and Systems*, vol. 2, July 2007, pp. 1–4.
[10] A. Makki, A. Siddig, M. Saad, J.R. Cavallaro, C.J. Bleakley, “Indoor Localization Using 802.11 Time Differences of Arrival,” *IEEE Trans. Instrum. Meas.*, V.65, No.3, pp.614-623, 2016.

[11] P. Biswas, H. Aghajan, and Y. Ye, “Integration of angle of arrival information for multimodal sensor network localization using semidefinite programming,” in *Proceedings of the 39th Asilomar Conference on Signals, Systems and Computers*, Oct. 2005.

[12] D. Niculescu and B. Nath, “Ad hoc positioning system (APS) using AOA,” in *Proceedings of the 22nd Annual Joint Conference of the IEEE Computer and Communications*, vol. 3, Mar. 2003, pp. 1734–1743.

[13] P. Bahl and V. Padmanabhan, “RADAR: an in-building RF-based user location and tracking system,” in *Proceedings of the 19th Annual Joint Conference of the IEEE Computer and Communications Societies*, vol. 2, 2000, pp. 775–784.

[14] C. Wu, Z. Yang, Y. Liu and W. Xi, "WILL: Wireless Indoor Localization without Site Survey," *IEEE Trans. Parallel Distrib. Syst.*, vol. 24, no. 4, pp. 839-848, April 2013.

[15] S. He and S.-H. G. Chan, "Wi-Fi Fingerprint-Based Indoor Positioning: Recent Advances and Comparisons," *Commun. Surveys Tuts.*, vol. 18, no. 1, pp. 466-490, 2016.

[16] Y. Lin, P. Tseng and Y. Chan, "Super-resolution-aided Positioning Fingerprinting based on Channel Impulse Response Measurement," in *Wireless Communications and Networking Conference (WCNC), 2017 IEEE*, San Francisco, CA, USA, 2017.

[17] X. Shen, K. Xu, X. Sun, J. Wu and J. Lin, "Optimized indoor wireless propagation model in WiFi-RoF network architecture for RSS-based localization in the Internet of Things," in *Microwave Photonics, 2011 International Topical Meeting on & Microwave Photonics Conference, 2011 Asia-Pacific, MWP/APMP*, Singapore, Singapore, 2011.

[18] A. H. Salamah, M. Tamazin, M. A. Sharkas and M. Khedr, "An enhanced WiFi indoor localization system based on machine learning," in *Indoor Positioning and Indoor Navigation (IPIN), 2016 International Conference on*, Alcala de Henares, Spain, 2016.

[19] X. Wang, L. Gao, S. Mao and S. Pandey, "CSI-Based Fingerprinting for Indoor Localization: A Deep Learning Approach," *IEEE Trans. Veh. Technol.*, vol. 66, no. 1, pp. 763-776, 2017.

[20] P.-H. Tseng, Y.-C. Chan, Y.-J. Lin, D.-B. Lin, N. Wu, T.-M. Wang, “Ray-Tracing-Assisted Fingerprinting Based on Channel Impulse Response Measurement for Indoor Positioning,” *IEEE Trans. Instrum. Meas.*, Vol. 66, No.5, pp.1032-1045, 2017.

[21] C. Shih, L. Chen, G. Chen, E. H. Wu and M. Jin, "Intelligent radio map management for future WLAN indoor location fingerprinting," in *Wireless Communications and Networking Conference (WCNC), 2012 IEEE*, Shanghai, China, 2012.

[22] Y. Chen, Q. Yang, J. Yin, and X. Chai, “Power-efficient access point selection for indoor location estimation*,” IEEE Trans. Knowl. Data Eng.*, vol. 18, no. 7, pp. 877–888, July 2006.

[23] A. Kushki, K. Plataniotis, and A. Venetsanopoulos, “Kernel-based positioning in wireless local area networks,” *IEEE Trans. Mobile Comput.*, vol. 6, no. 6, pp. 689–705, June 2007.

[24] C. Feng, W. Au, S. Valaee, and Z. Tan, “Received-signal-strength-based indoor positioning using compressive sensing,” *IEEE Trans. Mobile Comput.*, vol. 11, no. 12, pp. 1983–1993, Dec. 2012.

[25] V. Hodge and J. Austin, “A survey of outlier detection methodologies,” *Artificial Intelligence Review*, vol. 22, no. 2, pp. 85–126, Oct. 2004.

[26] Y. C. Chen, W. C. Sun, and J. C. Juang, “Outlier detection technique for RSS-based localization problems in wireless sensor networks,” in *Proceedings of SICE Annual Conference*, Aug. 2010, pp. 657–662.

[27] A. Khalajmehrabadi, N. Gatsis, D. Pack, and D. Akopian, “A joint indoor WLAN localization and outlier detection scheme using LASSO and Elastic-Net optimization techniques,” *IEEE Trans. Mobile Comput.*, vol. 16, no.8, pp. 2079 - 2092, 2017.

[28] J. A. Morales, D. Akopian, and S. Agaian, “Faulty measurements impact on wireless local area network positioning performance,” *IET Radar, Sonar Navig.*, vol. 9, no. 5, pp. 501–508, Aug. 2015.

[29] J. A. Morales, D. Akopian, and S. Agaian, “Mitigating anomalous measurements for indoor wireless local area network positioning,” *IET Radar, Sonar Navig*., vol. 10, no. 7, pp. 1220-1227, 2016.

[30] Tamosoft LTD, "Wireless Site Survey Software for 802.11 a/b/g/n/ac WLANs - Tamograph," Tamosoft, [Online]. Available: http://www.tamos.com/products/wifi-site-survey/. [Accessed June 2017].

[31] Iris Networks, "WLAN Site Survey Equipment," Iris Networks, [Online]. Available: http://irisnetworks.co.uk/terrawave/wlan-site-survey-equipment/. [Accessed June 2017].

[32] Berkeley Varitronics Systems, Wireless Detection. Yellowjacket-BANG Wi-Fi Analyzer. https://www.bvsystems.com/product/yellowjacket-bang-wi-fi-analyzer/ [Accessed August 2017].

[33] Wireshark, "CaptureSetup/WLAN - The Wireshark Wiki," Wireshark, [Online]. Available: https://wiki.wireshark.org/CaptureSetup/WLAN. [Accessed May 2017].

[34] C. Laoudias, M. P. Michaelides, C. G. Panayiotou, “Fault Detection and Mitigation in WLAN RSS Fingerprint-based Positioning,” in *Indoor Positioning and Indoor Navigation (IPIN), 2011 International Conference on*, Guimaraes, Portugal, Sept. 2011.

[35] A. Kushki, K. N. Plataniotis, A. N. Venetsanopoulos, “Kernel-Based Positioning in Wireless Local Area Networks,” *IEEE Trans. Mobile Comput.*, vol. 6, no. 6, pp. 689-705, Jun. 2007.

[36] A. Khalajmehrabadi, N. Gatsis, and D. Akopian, “Structured group sparsity: A novel indoor WLAN localization, outlier detection, and radio map interpolation scheme,” *IEEE Trans. Veh. Technol.*, vol. 66, no. 7, pp. 6498–6510, 2017.

[37] J. Talvitie, M. Renfors, and E. S. Lohan, “Distance-based interpolation and extrapolation methods for RSS-based localization with indoor wireless signals,” *IEEE Trans. Veh. Technol.*, vol. 64, no. 4, pp. 1340–1353, Apr. 2015.

[38] J. Talvitie, E. S. Lohan, M. Renfors, “The effect of coverage gaps and measurement inaccuracies in fingerprinting based indoor localization,” in *Localization and GNSS (ICL-GNSS), 2014 International Conference on*, Helsinki, Finland, 2014.

[39] Wireshark, *Wireshark - Go Deep*, [Online]. Available: https://www.wireshark.org/. [Accessed November 2016].

[40] Aircrack-ng, *Aircrack-ng*, [Online]. Available: http://aircrack-ng.org/. [Accessed December 2016].

[41] Aircrack-ng, *compatibility_drivers,* [Online]. Available: http://www.aircrack-ng.org/doku.php?id=compatibility_drivers#which_is_the_best_card_to_buy. [Accessed 2017].

[42] *IEEE Standard 802.11-2012 - Part 11: Wireless LAN Medium Access Control (MAC) and Physical Layer (PHY) Specifications,* IEEE Standards Association, The Institute of Electrical and Electronics Engineers, Inc, 2012.

[43] National Instruments. *Software Defined Radio*. [Online]. Available: http://www.ni.com/en-us/shop/select/software-defined-radio-device. [Accessed August 2017].

[44] Acrylic WiFi. *Free WiFi scanner and channel scanner for windows,* [Online]. Available: https://www.acrylicwifi.com/en/wlan-software/wlan-scanner-acrylic-wifi-free/. [Accessed December 2016].

[45] Jam Software, *HeavyLoad - Free Stress Tool for Your PC. Jam Software*, [Online]. Available: https://www.jam-software.com/heavyload/. [Accessed December 2016].

[46] Ubuntu Kernel Team, *stress-ng*, [Online]. Available: http://kernel.ubuntu.com/~cking/stress-ng/. [Accessed December 2016].

[47] H. Zou, B. Huang, X. Lu, H. Jiang, L. Xie, “A robust indoor positioning system based on the Procrustes analysis and weighted extreme learning machine,” *IEEE Trans. Wireless Commun.*, vol.15, no.2, pp. 1252-1266, Feb. 2016.

[48] S-H Fang, C-H Wang, “A novel fused positioning feature for handling heterogeneous hardware problem,” *IEEE Trans. Commun.*, Vol. 63, No.7, July 2015, pp. 2713-2723.

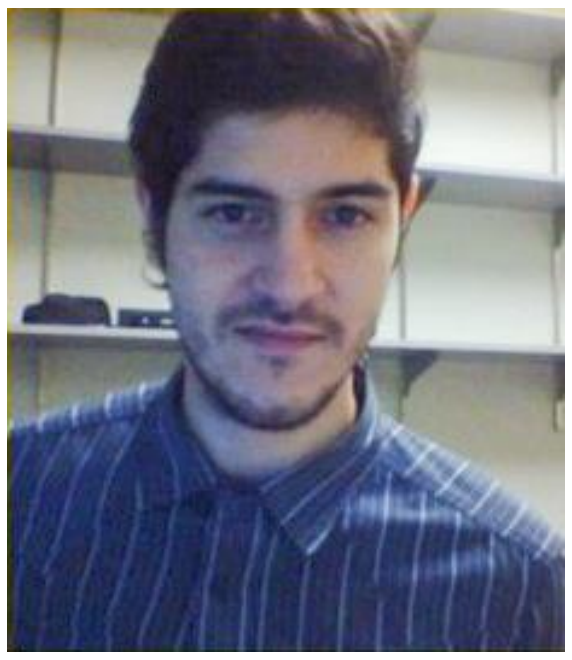

**Erick Schmidt** (S'17) received the B.Sc. degree (Hons.) in electronics and computer engineering from the Monterrey Institute of Technology and Higher Education, Monterrey, Mexico, in 2011, and the M.Sc. degree from the University of Texas at San Antonio, San Antonio, TX, USA, in 2015, where he is currently pursuing the Ph.D. degree in electrical engineering.

From 2011 to 2013, he was a Systems Engineer with Qualcomm Incorporated, San Diego, CA, USA. His current research interests include software-defined radio, indoor navigation, global navigation satellite system, and fast prototyping algorithms and accelerators for baseband communication systems.

Dr. Schmidt is a Student Member of the Institute of Navigation.

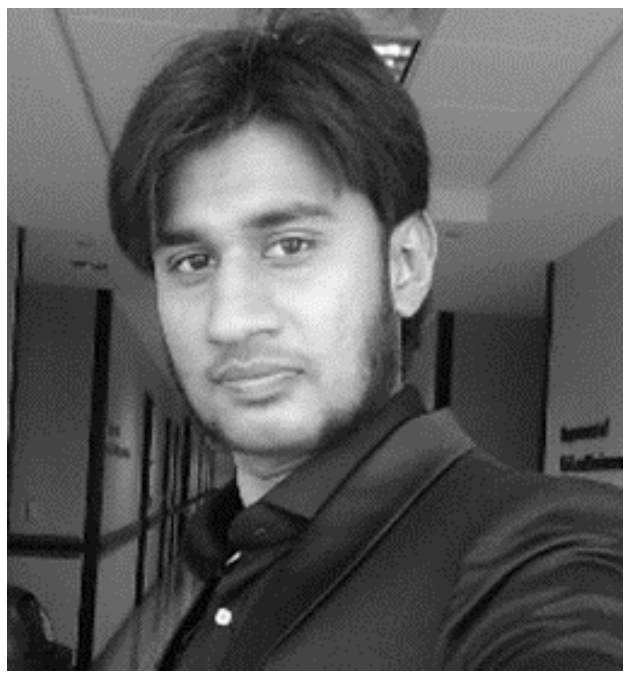

**Misbahuddin A. Mohammed** received the B.Sc. degree in electronics and communication engineering from Deccan College of Engineering and Technology, Osmania University, Hyderabad, India, in 2014, and the M.Sc. degree in electrical and electronics engineering from University of Texas at San Antonio, San Antonio, TX, USA, in 2017.

From 2014 to 2015, he had worked as a Technical Associate for Wipro ltd, Hyderabad, India. His interests are in the field of indoor navigation, data science and software engineering.

**David Akopian** (M'02–SM'04) received the Ph.D. degree from the Tampere University of Technology, Tampere, Finland, in 1997.

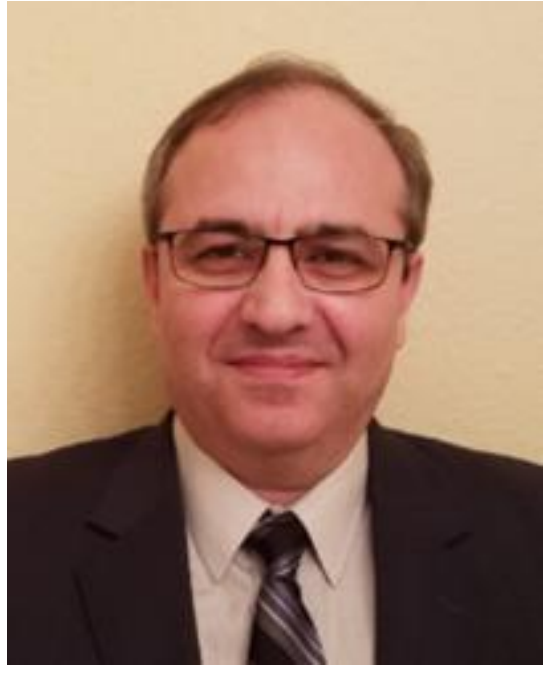

From 1993 to 1999, he was a Researcher and an Instructor at the Tampere University of Technology. From 1999 to 2003, he was a Senior Research Engineer and a Specialist with Nokia Corporation, Espoo, Finland. He is currently a Professor with the University of Texas at San Antonio, San Antonio, TX, USA. He has authored or co-authored over 30 patents and 140 publications. His current research interests include digital signal processing algorithms for communication and navigation receivers, positioning, dedicated hardware architectures and platforms for software-defined radio, and communication technologies for healthcare applications.

Dr. Akopian was elected as fellow of the U.S. National Academy of Inventors in 2016. He served in organizing and program committees of many IEEE conferences and co-chairs an annual conference on multimedia and mobile devices. His research has been supported by the National Science Foundation, the National Institutes of Health, USAF, US Navy, and Texas foundations.